\pdfoutput=1

\documentclass[11pt]{article}

\usepackage[final]{acl}

\usepackage{times}
\usepackage{latexsym}

\usepackage[T1]{fontenc}

\usepackage[utf8]{inputenc}

\usepackage{microtype}

\usepackage{inconsolata}

\usepackage{graphicx}
\usepackage{comment}
\usepackage{amsmath}
\usepackage{hyperref}       
\usepackage{url}            
\usepackage{booktabs}       
\usepackage{amsmath}        
\usepackage{amsfonts}       
\usepackage{amssymb}
\usepackage{amsthm}
\usepackage{nicefrac}       
\usepackage{microtype}      
\usepackage{xcolor}         
\usepackage{xspace}         

\newcommand{\given}{\,|\,}
\newcommand{\rank}{\textrm{rank}}

\newtheorem{theorem}{Theorem}[section]
\newtheorem{corollary}{Corollary}[theorem]

\newcommand{\GLIMPSEpoint}{$\textrm{GLIMPSE}_{\textit{point}}$\xspace}
\title{Efficient Pointwise-Pairwise Learning-to-Rank for News Recommendation}

\author{%
  \textbf{Nithish Kannen$^\star$$^\ddagger$\textsuperscript{1}},
  \textbf{Yao Ma$^\star$\textsuperscript{1}},
  \textbf{Gerrit J.J. van den Burg\textsuperscript{1}},
  \textbf{Jean Baptiste Faddoul$^\dagger$\textsuperscript{2}}
  \\
  \textsuperscript{1}Amazon,
  \textsuperscript{2}Zalando SE
  \\
  \texttt{nithishkannen@gmail.com},\\ \texttt{\{yaoom, gvdburg\}@amazon.com}, \\\texttt{jean.baptiste.faddoul@zalando.de}
  }

\begin{document}

\maketitle
\def\thefootnote{$\star$}\footnotetext{Equal contribution.}
\def\thefootnote{$\dagger$}\footnotetext{Work done at Amazon.}
\def\thefootnote{$\ddagger$}\footnotetext{This author is currently affiliated with Google DeepMind.}

\begin{abstract}

News recommendation is a challenging task that involves personalization based on the interaction history and preferences of each user. 
Recent works have leveraged the power of pretrained language models (PLMs) to directly rank news items by using inference approaches that predominately fall into three categories: pointwise, pairwise, and listwise learning-to-rank. While pointwise methods offer linear inference complexity, they fail to capture crucial comparative information between items that is more effective for ranking tasks. Conversely, pairwise and listwise approaches excel at incorporating these comparisons but suffer from practical limitations:  pairwise approaches are either computationally expensive or lack theoretical guarantees, and listwise methods often perform poorly in practice.
In this paper, we propose a novel framework for PLM-based news recommendation that integrates both pointwise relevance prediction and pairwise comparisons in a scalable manner. We present a rigorous theoretical analysis of our framework, establishing conditions under which our approach guarantees improved performance.  Extensive experiments show that our approach outperforms the state-of-the-art methods on the MIND and Adressa news recommendation datasets.

\end{abstract}

\section{Introduction}
\label{sec:introduction}

Online news services have become important platforms for a large population of users to stay informed. A massive number of news articles are generated and posted online every day, making it all the more important to personalize news recommendation for the users. The text-rich nature of news articles makes them particularly well-suited for the application of pre-trained language models (PLMs) \citep{wu2021empowering}. Despite the impressive advancements in large language models (LLMs), practical constraints in real-world news recommendation systems necessitate the use of more computationally efficient, and often smaller, PLMs.

One way to approach news recommendation using PLMs is through pointwise ranking. This method predicts a relevance score for each candidate based on the user’s previously clicked items. While this approach is scalable, the candidate scores are obtained \textit{independently} without comparing the relative usefulness of a candidate with regards to its competitors. Naturally, the task of recommendation requires \textit{comparing} candidates with each other. This intuition is supported by empirical results that demonstrate pairwise/listwise approaches outperform pointwise ones \citep{book}. Theoretically, listwise approaches are expected to perform the best as ranking is a list-level task \citep{10.1145/1273496.1273513, Liu2023GenerativeFN}. However, as shown by \citet{qin2023large}, listwise recommendation performs poorly when paired with PLMs for two main reasons: 1) PLMs often generate conflicting or useless outputs when provided large quantities of information, and 2) PLMs have a fixed maximum input prompt length that may be exceeded while attempting to encode all available candidate items. On the other hand, pairwise approaches leverage the comparison between items without the downsides of listwise approaches described above. However, they come with higher computational complexity compared to pointwise approaches. 

There is a body of relevant work on pairwise approaches for text-based recommendation tasks: \citet{qin2023large} propose \textit{pairwise prompting} for ranking with PLMs by performing bubble-sort at inference time. \citet{pradeep2021expandomonoduo} propose a two-stage approach that first ranks using pointwise scores and subsequently modifies the rankings by computing aggregation scores from pairwise comparisons. However, both these methods are computationally expensive and scale poorly in practice due to the requirement for $O(n^2)$ comparisons. To optimize pairwise approaches, \citet{Gienapp_2022} propose to sparsify the number of pairwise comparisons by random and skip-window sampling of pairs. While this improves scalability, it only compares several random aggregation strategies without theoretical guarantees. 

In this work we propose a theoretically-guaranteed method that efficiently performs pairwise ranking. We summarize our main contributions as follows:
\begin{itemize}
    \item A multi-task model jointly trained to perform both pointwise and pairwise predictions. We use a text-to-text approach where both are treated as classification tasks (prediction of pre-defined target words), hence aligning the tasks with the PLM pre-training objective.
    \item An efficient inference strategy in which we initialise with a ranking obtained from pointwise model scores and perform Right-To-Left (RTL) passes for pairwise reranking. That is, adjacently ranked elements are compared using the pairwise functionality starting from the rightmost and until the leftmost position. In this way, we perform only $n - 1$ pairwise comparisons per RTL pass.
    \item A theoretical framework for this approach based on Markov chains. We derive testable conditions that can be used to verify the strategy is beneficial for a given ranking metric.
    \item Extensive experiments that show our approach outperforms the state of the art on the MIND and Adressa news recommendation datasets.
\end{itemize}

We emphasize that although our focus is on news recommendation, our proposed algorithm is directly applicable to any text-based task.

\section{The Proposed Algorithm: GLIMPSE}

The task of news recommendation in this paper refers to ranking a set of candidate news items $X = \{x_1, x_2, x_3, \ldots, x_K\}$ given the user's click history. The goal is to rank an item higher than others if it is preferred by the user. This goal implies both a pointwise relevance prediction task and a pairwise preference task: a user click on an item should be treated not only as an absolute judgment for relevancy, but also as a preference judgement. Motivated by this, we propose the  \textbf{GLIMPSE} (\textbf{G}enerative Tup\textbf{L}e-W\textbf{I}se Pro\textbf{MP}ting for New\textbf{S} recomm\textbf{E}ndation) algorithm for optimizing both objectives simultaneously. GLIMPSE is a general method for recommendation problems and works with any generative model. 

\begin{figure*}[!t]
  \centering
  \tiny 
  \includegraphics[keepaspectratio,width=0.75\textwidth]{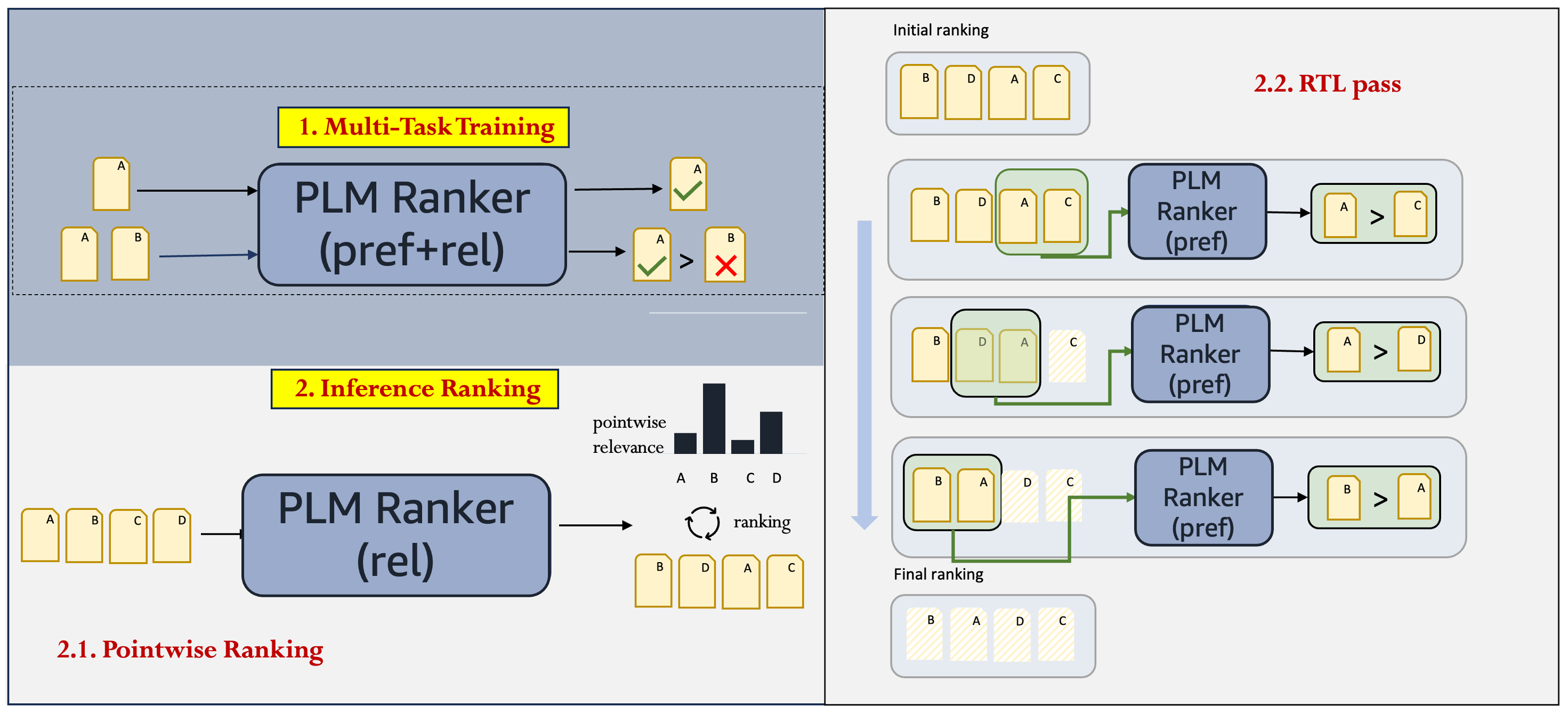}
  \caption{An illustration of the proposed framework. GLIMPSE consists of a multi-task training approach where the PLM is fine-tuned by considering both the relevance prediction and the pairwise preference tasks. During inference, the relevance predictions are used to obtain an initial pointwise ranking, which is subsequently improved by performing one or more right-to-left (RTL) passes using pairwise comparisons.
  }
  \label{fig:main}
\end{figure*}

\if We propose \textbf{GLIMPSE} (\textbf{G}enerative Tup\textbf{L}e-W\textbf{I}se Pro\textbf{MP}ting for New\textbf{S} recomm\textbf{E}ndation). We use a fully text-to-text based approach with an encoder-decoder transformer model. We train a a multi-task model to perform both pointwise and pairwise ranking. Note that during inference, pointwise score prediction only takes $O(N)$ model calls, whereas a pairwise approach can take atmost $O(N^2)$ and at best $O(N log(N))$  comparisons, making it an impractical ranking solution. Our inference strategy finds a middle ground by initialising with pointwise ranking and then using unidirectional (right-to-left) adjacent element swaps with pairwise comparisons as shown in Figure \ref{fig:main}. We discuss more details about the of multitask training and inference strategy in the this sections. \fi

We first describe the multi-task fine-tuning method of a generative model that aligns both the relevance prediction objective and the preference prediction objective into a single text generation task. By performing multi-task fine-tuning, we obtain a single model that can be used as a predictor for both objectives at the same time. To  combine the tasks into a complete ranking method, we subsequently propose a novel inference strategy by initialising with pointwise relevance ranking and using unidirectional (right-to-left) adjacent element swaps with pairwise preference comparisons to obtain a final ranking (illustration in Figure~\ref{fig:main}). 


\subsection{Multi-task Fine-tuning for Ranking \label{sec:multitask}}


The proposed multitask fine-tuning strategy is motivated by the fact that jointly optimizing ranking and classification objectives has been shown to achieve better performance than using ranking-only or classification-only objectives, especially when observations are limited \cite{kdd2010:Combined}. The strategy aligns the two objectives into a single one, in our case through text generation.

\paragraph{Pointwise Relevance Prediction (Rel).} The first task we consider is a classification task where the goal is to predict the relevance of a candidate item given the user history. This task looks at a \textit{single} candidate item at a time. It involves predicting $y_{ij} \in \{0, 1\}$ for a candidate item $x_i$ conditioned on user's click history $\mathcal{H}_j$. In practice, this is a classification task to classify if a candidate $x_i$ is relevant ($y_{ij}=1$) or not relevant ($y_{ij}=0$) given history $\mathcal{H}_j$. To fine-tune the generative model on this task, we prompt the input text sequence to essentially answer the question, \textit{Is the item a suitable recommendation for the user?} After fine-tuning, we obtain a relevance prediction function $\text{Rel}$ which gives the probability that an item is relevant for a user with history $\mathcal{H}_j$, i.e.,
\begin{equation}
\begin{array}{lll}
    \text{Rel}(x_i \given \mathcal{H}_j) := P({y}_{ij}=1|x_i,\mathcal{H}_j).
\end{array}
\end{equation}
In practice, we use the probability of the positive target token as the probability of relevance.

\paragraph{Pairwise Preference Prediction (Pref).} In the second task, the model is asked which candidate it prefers when it is provided a pair of options. More specifically, given two candidate items $x_1$ and $x_2$ and user history $\mathcal{H}_j$, the model predicts the probability of the preference of one candidate over the other. This too can be framed as a text generation task, through the question: \textit{Given item 1 and item 2, which is a more suitable recommendation for the user?} After fine-tuning, we obtain a prediction of the probability that an item is preferred over another for a specific user, i.e.,
\begin{equation}
    \text{Pref}((x_i, x_j) | \mathcal{H}_j) := P(x_i>x_j|\mathcal{H}_j).
\end{equation}
Similar to the relevance prediction task, the probability of preferring one candidate over another is based on the probability of predicting the corresponding target tokens. To obtain $P(x_i > x_j \given \mathcal{H}_j)$ from these token probabilities we leverage the Bradley-Terry realization \cite{Bradley1952RankAO}. Writing $\delta_i$ and $\delta_j$ for the probability of predicting the target words \emph{Candidate A} and \emph{Candidate B} respectively for $x_i$ and $x_j$, we have:
\begin{equation}
\label{eq:pairwise}
    P(x_i > x_j) =  \frac{e^{\delta_i}}{e^{\delta_i} +e^{\delta_j} }.
\end{equation}

\paragraph{Multi-task training.} Similar to previous work \cite{su-etal-2022-multi}, we use discrete task-specific prompts to differentiate between the tasks. We cast both tasks into text-generation problems with task prompts and pre-defined target words. In the multi-task fine-tuning stage, the training sample is represented as a tuple $d = (s_{t},x,y,\mathcal{H})$. Here $s_t$ refers to the task-prompt with $t \in \{\text{Rel}, \text{Pref}\}$, and $x$ refers to the input candidates. Note that  $x$ corresponds to a single item in the \text{Rel} task and a pair of items in the \text{Pref} task. Table~\ref{tab:prompt_table} in Appendix~\ref{app:expdetails} shows examples of task-prompts and target words. For the relevance task we sample an equal number of positive (clicks) and negative (no-click) samples, while for the preference task we construct a sample by picking one positive and one negative item from the same impression. An impression refers to user activity information containing user click history, including positives and negatives. The set of positives and negatives form the candidate set of an impression. During training, we shuffle and mix data points from different tasks for multi-task fine-tuning \cite{su-etal-2022-multi}. Thus the model is trained by maximizing the likelihood objective defined as
\begin{equation}
\label{eq:LED}
\mathcal{L}_{\theta} = -\sum_{d}\sum_{l=1}^{|y|}\log\; p_{\theta}(y^l|y^{<l};s_t,x,\mathcal{H}),
\end{equation}
where $|y|$ denotes the length of the target sequence $y$, $y^l$, $y^{<l}$ are the $l$-th token and tokens before $l$, and $\theta$ denotes the model parameters. 







\subsection{Aggregated Ranking Inference}

The multi-task fine-tuned model can both predict a pointwise relevance score and perform pairwise preference comparison. Here we introduce a novel inference approach that incorporates both of these capabilities and leverages the different benefits of pointwise and pairwise predictions for learning to rank.

\paragraph{Rank Aggregation Strategy.} We initialize the ranked list by predicting the relevance score $\text{Rel}(x^i)$ for each candidate item $x_i$ and sorting by this score in descending order. Next we perform local refinement by applying the pairwise preference scores to pairs of items in this sorted list. Specifically, we perform $m$ RTL passes on the top-$k$ ranked items in the sorted list. An RTL pass is defined as a single pass in the right to left direction where adjacent items are compared using the $\text{Pref}(\cdot,\cdot)$ function starting from the rightmost to the leftmost item. Two adjacent items $x^i$ and $x^{i+1}$ are swapped if $\text{Pref}(x^i,x^{i+1}) < \text{Pref}(x^{i+1},x^i)$. The number of RTL passes $m$ and the top-$k$ items considered in an RTL pass are hyperparameters of our inference approach.

The proposed RTL rank aggregation algorithm is guaranteed to achieve better ranking performance under certain conditions compared to the pointwise relevance ranking strategy (see Section~\ref{sec:theory}). It is clear that one RTL pass of top-$k$ elements consists of $k - 1$ comparisons, as shown in Figure \ref{fig:main}. Note that when $k = K$, it leads to the classic Bubblesort algorithm \citep{friend1956sorting} except the sorting is done in a stochastic manner. The pointwise score prediction only takes $O(K)$ model calls, whereas a pairwise approach can take at most $O(K^2)$ and at best $O(K \log K)$  comparisons with Quicksort, making it an impractical ranking solution. Our approach provides a practical middle ground with $O(K)$ complexity.



\section{Theoretical Analysis}
\label{sec:theory}
In this section, we provide the theoretical analysis for the proposed RTL rank aggregation strategy. Let $\kappa$ be a permutation of the indices $\{1, \ldots, K\}$ of candidate items $\{x_1,\ldots,x_K\}$. We denote $\rank(x_i|\kappa)$ as the rank of item $x_i$ in the permutation $\kappa$ and write $r^*(x_i|\mathcal{H}) \in \{0, 1\}$ for the ground-truth relevance of item $x_i$ with history $\mathcal{H}$.
We can then consider additive ranking metrics of the form:
\begin{equation}
    \label{eq:metric_definition}
    \Delta(\kappa \given \mathcal{H}, r^*) = \sum_{i=1}^K \lambda(\rank(x_i \given \kappa)) \cdot r^*(x_i|\mathcal{H}),
\end{equation}
where $\lambda$ is a function over the rank \citep{agarwal2019general}. For example, the well-known discounted cumulative gain (DCG) metric \citep{jarvelin2002cumulated} corresponds to $\lambda(v) = 1/\log(1 + v)$.  
For a stochastic algorithm $\mathcal{A}$ we are interested in the expectation of the ranking metric over the possible permutations $\mathbf{\kappa}$, which we can write as $
    \mathbb{E}_{\mathcal{A}}(\Delta) = \sum_{\kappa} p(\kappa) \Delta(\kappa \given \mathcal{H}, r^*)$,
with $p(\kappa)$ the probability of obtaining the ranking $\kappa$ from the algorithm for a particular user.

Because the relevance $r^*(\cdot)$ is a binary value (e.g., click or no click), distinct permutations $\kappa$ can give rise to the same value of the ranking metric. Let $\mathbf{z} \in \{0, 1\}^K$ denote the binary vector of user relevance of items in ranking $\kappa$, such that $\mathbf{z} = [r^*(x_i|\mathcal{H}) : i \in \kappa]$. In this setting, the ranking metric becomes $
    \Delta(\mathbf{z} \given \mathcal{H}) = \sum_{k=1}^K \lambda(k) z_k$,
with $z_k$ the $k$-th element of $\mathbf{z}$. By extension, we can write the expectation over the metric in terms of $\mathbf{z}$ as well, $
    \mathbb{E}_{\mathcal{A}}(\Delta) = \sum_{\mathbf{z}} p(\mathbf{z}) \Delta(\mathbf{z} \given \mathcal{H})$,
where the summation is over all possible values of $\mathbf{z}$.

We can refine the expression for the expected ranking metric by considering our proposed two-stage inference approach consisting of an initial ranking of the candidates based on relevance prediction, and a subsequent refinement of that ranking through pairwise preferences. Let $\boldsymbol{\pi} \in \mathbb{R}^{2^K}$ denote the probability vector with elements $p(\mathbf{z})$ for all $\mathbf{z} \in \{0, 1\}^K$ that reflects the probability of obtaining $\mathbf{z}$ from pointwise inference. Then, the ranking refinement from pairwise comparisons can be modeled as a Markov chain with transition matrix $\mathbf{T} \in \mathbb{R}^{2^K \times 2^K}$, such that the product $\boldsymbol{\pi}' = \boldsymbol{\pi}\mathbf{T}$ is the probability vector of ``states'' after the transition induced by pairwise reranking. Specifically, the elements $T_{i,j}$ of the transition matrix $\mathbf{T}$ represent the probability of transitioning from a state $\mathbf{z}^i$ to a state $\mathbf{z}^j$. With a mild abuse of notation we can write $\mathbf{T}_{\mathbf{z}'}$ for the column of $\mathbf{T}$ corresponding to an outcome state $\mathbf{z}'$. Then, the expected value of the ranking metric after our two-stage inference process becomes
\begin{equation}
\label{equation:expectedrankingmetric}
    \mathbb{E}_{\mathcal{A}}(\Delta) = \sum_{\mathbf{z}'} \boldsymbol{\pi} \mathbf{T}_{\mathbf{z}'} \Delta(\mathbf{z}' \given \mathcal{H}).
\end{equation}
Repeated application of the pairwise refinement (i.e., multiple RTL passes) can be captured through typical Markov chain notation,
    $\mathbb{E}_{\mathcal{A}}(\Delta) = \sum_{\mathbf{z}'} \boldsymbol{\pi} \mathbf{T}^{\alpha} \mathbf{T}_{\mathbf{z}'} \Delta(\mathbf{z}' \given \mathcal{H})$,
for $\alpha \in \mathbb{Z}^+$ a non-negative integer. We next analyze the transition matrix to understand the conditions under which the RTL passes are beneficial.

\subsection{Transition Matrix}
The transition matrix $\mathbf{T}$ is parameterized by the output of the pairwise preference predictions. Since the model predicts the preference over any pair of items, we can define the probability of swapping or not swapping items with distinct user relevance values. Formally, for any pair of items $(i,j)$, we can define the following probabilities:
\begin{equation}
\label{equation:swap-no-swap}
    \begin{array}{l}
        \mu := P(x_i > x_j \given r^*(x_i) = 1, r^*(x_j) = 0)  \\
        \nu := P(x_i > x_j \given r^*(x_i) = 0, r^*(x_j) = 1).  \\
    \end{array}
\end{equation}
Thus, $\mu$ is the probability that the model predicts that $x_i$ is preferred over $x_j$ and this \emph{is} the case in the ground truth, while $\nu$ is the probability that the model predicts that $x_i$ is preferred over $x_j$ even though this \emph{is not} the case in the ground truth. The converse probabilities $1 - \nu$ and $1 - \mu$ can be defined analogously.

For a specific inference strategy, we can derive the corresponding transition matrix using the number of \textbf{swap} and \textbf{no-swap} steps needed to transit between states. Mathematically, any element of the transition matrix $\mathbf{T}$ can be expressed in the following form $
    \mu^{\alpha_1}(1-\mu)^{\alpha_2}\nu^{\alpha_3}(1-\nu)^{\alpha_4}$.
where $\alpha_1,\alpha_2,\alpha_3,\alpha_4$ are the numbers of swap and no-swap steps needed to transit from a state to another. 

Now, we show the transition matrix for the proposed RTL rank aggregation strategy. For simplicity we assume $\forall \mathbf{z},\|\mathbf{z}\|_1=1$, which indicates there is a single relevant item. In this case, it is clear that the state space consists of $K$ states $\mathbf{z}^1,...,\mathbf{z}^K$, where $\mathbf{z}^1=[1,0,...,0],\mathbf{z}^2=[0,1,0,...,0],...,\mathbf{z}^K=[0,...,0,1]$. Note that the Mean Reciprocal Rank (MRR) metric can be written by using $\lambda(v)=1/v$ under this assumption. We denote this metric as $\Delta_{M}$ below. We can write the transition matrix $\mathbf{T}$ of the pairwise refinement, whose $i,j$-th element equals $P(\mathbf{z}^j|\mathbf{z}^i)$, as
\begin{equation}
\label{Equation:transitionMatrix}
    T_{i,j}= \left\{ 
    \begin{array}{ll}
    0 & (a) \\
    \mu & (b)\\
    (1-\mu)(1-\nu)\nu^{i-j} & (c) \\
    (1-\nu)\nu^{K-j} & (d) \\
    (1-\mu)\nu^{i-1} & (e) \\
    \nu^{K - 1} & (f),
    \end{array}\right.
\end{equation}
where 
    (a) $\textrm{ if } j > i + 1 \textrm{ and } 1 \leq i < K$,
    (b) $\textrm{ if } j = i + 1 \textrm{ and } 1 \leq i < K$,
    (c) $ \textrm{ if } 1 < j \leq i \textrm{ and } 1 < i < K$,
    (d) $\textrm{ if } 1 < j \leq i \textrm{ and } i = K$,
    (e) $\textrm{ if } j = 1 \textrm{ and } 1 \leq i < K$, and
    (f) $\textrm{ if } j = 1 \textrm{ and } i = K$.

\if By plugging in the transition matrix into (\ref{equation:expectedrankingmetric}), the expected value of the ranking metric can be explicitly evaluated. We first obtain:
\begin{align}
    \mathbb{E}_{\mathcal{A}}(\Delta) &= \sum_{j=1}^K \boldsymbol{\pi} \mathbf{T}_j \Delta(\mathbf{z}^j) = \sum_{j=1}^K \sum_{i=1}^{K} \mathbf{\pi}_i T_{i,j} \Delta(\mathbf{z}^j) = \sum_{i=1}^K \mathbf{\pi}_i \sum_{j=1}^{i+1} P(\mathbf{z}^j \given \mathbf{z}^i) \Delta(\mathbf{z}^j) \nonumber \\
     &= \sum_{i=1}^{K-1} \pi_i \mu \Delta(\mathbf{z}^{i+1}) + \sum_{i=2}^{K-1} \sum_{j=2}^i \pi_i (1 - \mu)(1 - \nu) \nu^{i - j} \Delta(\mathbf{z}^j) \nonumber \\
     &\quad + \sum_{j=2}^{K} \pi_K (1 - \nu)\nu^{K - j} \Delta(\mathbf{z}^j) + \sum_{i=1}^{K-1} \pi_i (1 - \mu) \nu^{i - 1} \Delta(\mathbf{z}^1) + \pi_K \nu^{K-1} \Delta(\mathbf{z}^1).
\end{align}
By collecting terms, this becomes
\begin{align}
    \mathbb{E}_{\mathcal{A}}(\Delta) &= \pi_1 \left( \mu \Delta(\mathbf{z}^2) + (1 - \mu)\Delta(\mathbf{z}^1) \right) \nonumber \\
    &\quad + \sum_{i=2}^{K-1} \pi_i \left( \mu \Delta(\mathbf{z}^{i+1}) + \sum_{j=2}^i (1-\mu)(1-\nu)\nu^{i-j} \Delta(\mathbf{z}^{j}) + (1 - \mu)\nu^{i-1}\Delta(\mathbf{z}^1) \right) \nonumber \\
    &\quad + \pi_K \left( \sum_{j=2}^{K-1} (1 - \nu)\nu^{K-j}\Delta(\mathbf{z}^j) + (1-\nu)\Delta(\mathbf{z}^K) + \nu^{K-1}\Delta(\mathbf{z}^1) \right).
\end{align}

Using this expression we can investigate the conditions under which the proposed RTL inference strategy is beneficial for the expected metric value. We consider three cases for the source state $\mathbf{z}^i$ to understand the reranking behavior in more detail.
\begin{enumerate}
    \item For $i=1$, the pointwise ranker gives the best ranking result, $\mathbf{z}^1$. The best outcome for the pairwise reranking would therefore be to maintain the status quo and not perform any swaps. Conditional on $\mathbf{z}^i = \mathbf{z}^1$ the expected metric value for this case is
    \begin{equation}
        \mathbb{E}_{\mathcal{A}}(\Delta \given \mathbf{z}^i = \mathbf{z}^1) = \mu \Delta(\mathbf{z}^2) + (1 - \mu) \Delta(\mathbf{z}^1).
    \end{equation}
    Thus, in order to achieve a better or equal metric score, we require that
    \begin{equation*}
        (1-\mu)\Delta(\mathbf{z}^1) + \mu\Delta(\mathbf{z}^2) \geq \Delta(\mathbf{z}^1),
    \end{equation*}
    which clearly holds only when $\mu=0$ (i.e., the probability of a ``bad swap'' is 0). When $\mu\neq 0$, we suffer a degradation in the metric value equal to $\mu (\Delta(\mathbf{z}^1) - \Delta(\mathbf{z}^2)) = \mu(\lambda(1) - \lambda(2))$, where $\lambda$ is a function over the rank (see eq.~\ref{eq:metric_definition}). For the Mean Reciprocal Rank (MRR) metric, this degradation equals $\mu/2$, whereas for the DCG metric it is $\mu (1/\log(2) - 1/\log(3)) \approx 0.53 \mu$.
    \item For $1 < i < K$ we obtain improvement conditional on being in state $\mathbf{z}^i$ if
    \begin{equation}
        \mu \Delta(\mathbf{z}^{i+1}) + (1 - \mu)\nu^{i-1}\Delta(\mathbf{z}^i) + \sum_{j=2}^i (1 - \mu)(1 - \nu) \nu^{i-j} \Delta(\mathbf{z}^j) \geq \Delta(\mathbf{z}^i).
    \end{equation}
    By plugging in specific forms of the ranking metric, this expression can yield a tight inequality on $\mu$ and $\nu$ to ensure improvement from pairwise ranking (see \textcolor{red}{Appendix TODO}). In general however, we can derive a simple expression without specifying the metric through intuitive reasoning. Recall that $\mu$ is the probability of swapping $(1, 0)$ to $(0, 1)$ in $\mathbf{z}^i$ (a ``bad'' swap) and $\nu$ is the probability of swapping $(0, 1)$ to $(1, 0)$ (a ``good'' swap). Thus we obtain improvement if the probability of making a bad swap is smaller than the probability of not making a bad swap and then making a good swap, i.e., $\mu \leq (1 - \mu)\nu$. Because this reasoning makes no assumption about $\mathbf{z}$, it holds regardless of whether $||\mathbf{z}||_1 = 1$ or not. Rewriting as a condition on $\mu$ gives $\mu \leq \nu/(\nu + 1)$.
    \item If $i = K$ the pointwise ranker placed the clicked item at the end of the ranking. Thus, any pairwise reranking will not degrade the ranking metric. Extracting the $i = K$ portion of the expected ranking metric above, we find that we obtain improvement from pairwise reranking as long as:
    \begin{equation}
        \sum_{j=2}^{K-1} (1 - \nu) \nu^{K-j} \Delta(\mathbf{z}^j) + (1 - \nu)\Delta(\mathbf{z}^K) + \nu^{K-1}\Delta(\mathbf{z}^1) \geq \Delta(\mathbf{z}^K).
    \end{equation}
    Which is equivalent to
    \begin{equation}
        (1 - \nu) \sum_{j=2}^{K-1} \nu^{K-j} \Delta(\mathbf{z}^j) + \nu^{K-1} \Delta(\mathbf{z}^1) \geq \nu \Delta(\mathbf{z}^K)
    \end{equation}

\end{enumerate}
\textcolor{red}{TODO HERE: Below needs to be brought in line with the above}

In order to prove the improvement of MRR after applying the transition matrix, we analyze the MRR differences from each state $\mathbf{z}i$ separately. 
\begin{enumerate}
    \item For $i=1$, the pointwise ranker gives perfect ranking result who has $z_1=1, z_2=0, ..., z_K=0$. In order to achieve better or equal Mrr score, we have
    \begin{equation*}
        (1-\mu)\textrm{Mrr}(\mathbf{z}_1)+\mu\textrm{Mrr}(\mathbf{z}_2)\geq\textrm{Mrr}(\mathbf{z}_1),
    \end{equation*}
where the equality holds only when $\mu=0$. When $\mu\neq 0$, we suffer Mrr degradation of $\mu/2$.

    \item For $1<i<K$, we can write the general condition for achieving better Mrr score after pairwise refinement as:
    \begin{equation*}
        \mu\textrm{Mrr}(\mathbf{z}_{i+1})+(1-\mu)\nu^{i-1}\textrm{Mrr}(\mathbf{z}_1)+\sum_{j=2}^{i}(1-\mu)(1-\nu)\nu^{i-j}\textrm{Mrr}(\mathbf{z}_j)\geq \textrm{Mrr}(\mathbf{z}_i),
    \end{equation*}
    which can be expressed as after plugging in the definition of Mrr
    \begin{equation*}
        i\nu^{i-1}+(1-\nu)\sum_{j=2}^i\frac{i}{j}\nu^{i-j}\geq\frac{(1-\mu)i+1}{(i+1)(1-\mu)}.
    \end{equation*}
    After a re-arrangement, we can write the condition for achieving non-negative improvement as
    \begin{equation*}
        \mu\leq\frac{\Lambda_i}{\Lambda_i+1},
    \end{equation*}
    where $\Lambda_i=\frac{i(i+1)}{2}\nu^{i-1}+\frac{i(i+1)}{2*3}\nu^{i-2}+\frac{i(i+1)}{3*4}\nu^{i-3}+...+\frac{i(i+1)}{(i-1)*i}\nu$.
    As $\frac{\Lambda_i}{\Lambda_i+1}$ is a monotonic increasing function in terms of $i$ and $\nu$, we are guaranteed to have non-negative improvement when $\mu\leq\frac{3\nu}{3\nu+1},\forall i=2,...,K-1$.
    \item For $i=K$, we always gain non-negative improvement since the pointwise model gives the worst Mrr score. 
\end{enumerate}
\fi

With the defined transition matrix $\mathbf{T}$, we have the following theorem (proof in Appendix~\ref{app:proof}). 
\begin{theorem}
\label{the:main}
For any $\boldsymbol{\pi} = [p(\mathbf{z}^1),p(\mathbf{z}^2),..,p(\mathbf{z}^K)]$ obtained from a pointwise inference strategy, the RTL pairwise ranking refinement achieves positive gain in terms of the expected MRR, $\mathbb{E}_{A}(\Delta_{M})-\mathbb{E}_{\pi}(\Delta_{M})>0$, when the pairwise inference satisfies
\begin{equation}
\mu<\frac{p(\mathbf{z}^i)}{p(\mathbf{z}^{i+1})}\nu, \quad \forall i =1,...,K-1.
\end{equation}
\end{theorem}
\begin{corollary}
By applying the pairwise ranking refinement inference $\alpha$ times, the expected MRR achieves positive gain if
\begin{equation}
    \mu<\frac{\pi T^{\alpha}_{i}}{\pi T^{\alpha}_{i+1}}\nu, \qquad \forall i=1,\ldots,K,
\end{equation}
where $\pi T^{\alpha}_i$ is the $i$-th element of vector $\pi T^{\alpha}$.
\end{corollary}

Intuitively,  Theorem~\ref{the:main} shows that the proposed two-stage inference achieves better performance when the pairwise inference model outperforms the pointwise inference model. Here, we also provide the analysis for the state distribution $\boldsymbol{\pi}$ obtained from the pointwise inference model. We assume the relevancy score $s=\text{Rel}(\cdot)\in[0,1]$ predicted by the pointwise ranker follows two beta distributions, where the predicted score for positive and negative classes follow $\textrm{Beta}(\alpha_1,\beta_1)$ and $\textrm{Beta}(\alpha_2,\beta_2)$, respectively. For a given impression containing one positive item and $K-1$ negative items, the probability $p(\mathbf{z}^k)$ can be written as
\begin{equation}
\label{Equ:prior}
\begin{array}{l}
    p(\mathbf{z}^k)=\binom{K-1}{k-1}\int_{0}^{1}(1-F(u;\alpha_2,\beta_2))^{k-1}\\
    F(u;\alpha_2,\beta_2)^{K-k}f(u;\alpha_1,\beta_1)\mathrm{d}u,
\end{array}
\end{equation}
where $\binom{K-1}{k-1}$ denotes the binomial coefficient, and $f(\cdot)$ and $F(\cdot)$ are the probability density and cumulative distribution functions of the beta distribution. With the defined state probability, we have for all $k=1,...,K-1$, $
    \frac{p(\mathbf{z}^k)}{p(\mathbf{z}^{k+1})}\leq\frac{p(\mathbf{z}^1)}{p(\mathbf{z}^2)}$.
Together with Theorem~\ref{the:main}, we can rewrite the condition for obtaining positive gain after pairwise refinement as $\mu\leq\nu\frac{p(\mathbf{z}^1)}{p(\mathbf{z}^2)}$. Though a closed form condition is hard to achieve, the ratio $p(\mathbf{z}^1)/p(\mathbf{z}^2)$ can be estimated through a numerical simulation with Eq.~\ref{Equ:prior}.

\section{Experiments}
\label{sec:experiments}

\begin{table*}[tb]
\centering
\footnotesize
\begin{tabular}{lcccccccc}
\toprule
& \multicolumn{4}{c}{\textbf{MIND}} & \multicolumn{4}{c}{\textbf{Adressa}}  \\
\cmidrule(lr){2-5} \cmidrule(lr){6-9}  
\textbf{Model} & \textbf{AUC} & \textbf{MRR} & \textbf{nDCG@5} & \textbf{nDCG@10} & \textbf{AUC} & \textbf{MRR} & \textbf{nDCG@5} & \textbf{nDCG@10} \\
\midrule
BERT-NPA$^{\diamondsuit}$ & 67.56 & 31.94 & 35.34 & 41.73 & 56.21 & 29.44 & 26.98 & 32.67\\
BERT-LSTUR$^{\diamondsuit}$ & 68.28 & 32.58 & 35.99 & 42.32 & 56.76 & 29.87 & 27.34 & 32.90\\
BERT-NRMS$^{\diamondsuit}$ & 68.60 & 32.97 & 36.55 & 42.78 & 56.95 & 30.08 & 27.89 & 32.92\\
DIGAT & 68.77 & 33.46 & 37.14 & 43.39 & 57.13 & 30.18 & 27.95 & 32.90 \\ 
HDNR & 68.23 & 32.61 & 36.10 & 42.29 & 57.43 & 30.09 & 28.34 & 34.11 \\ 
Prompt4NR$^{\spadesuit}$ & 68.48 & 33.29 & 37.12 &	43.25 & 61.67 & 30.32 & 28.98 & 36.33 \\
UniTRec & 68.59	& 33.76	& 37.63 & 43.74 & 62.29 & \textbf{31.90} & 29.43 & 36.90 \\
\midrule
\GLIMPSEpoint & 68.18 &	33.56 &	37.27 & 43.48 & 57.82 & 30.38 & 26.00 & 33.14\\
\quad w/ pair (1 pass, top 2) & 68.88 &	33.70 & 37.38 &	43.59 & \textbf{67.24} & 31.84 & \textbf{30.98} & \textbf{39.33} \\
\quad w/ pair (1 pass, top 3) & 68.92 & 33.71 &	37.39 &	43.60 & 63.15 & 27.90 & 27.90 & 34.95 \\ 
\quad w/ pair (1 pass, top 5) & 68.97 &	33.73 &	37.41 &	43.62 & 67.23 & 30.65 & 30.13 & 38.47 \\
\quad w/ pair (2 pass, top 5) & \textbf{69.14} & \textbf{33.88} & \textbf{37.66} & \textbf{43.88} & 67.14 & 30.86 & 30.25 & 38.60 \\ 
\bottomrule
\end{tabular}
\caption{Results on the MIND and Adressa datasets compared to baseline methods. All results on Adressa are reproduced using publicly-available code. For MIND, $\spadesuit$ means that we reproduce the scores using the publicly-available code, and $\diamondsuit$ indicates that results are as reported in \citet{Zhang_2023}. For other methods results are extracted from the respective papers. \label{tab:combined_results}}
\end{table*}

In this section we conduct experiments to empirically validate our approach. We first describe a comparison with existing works. Next, we describe an experiment on various inference strategies, followed by an ablation study to better understand the effect of each component of our approach. Additional details about the experimental setup can be found in Appendix~\ref{app:expdetails}.
\subsection{Setup}
Our experiments focus on two commonly-used datasets for news recommendation, MIND \citep{wu-etal-2020-mind} and Adressa \citep{gulla2017adressa}. The Microsoft News Dataset (MIND) contains information regarding user sessions on a news aggregation website. User sessions are organized in ``impressions'', where an impression consists of an ordered list of news articles and the user click behavior for those articles. To support fair comparison to previous work \citep{Zhang_2023, bi-etal-2022-mtrec,qi2021hierec}, we use the MIND-Small subset of MIND for training, which consists of click-behavior for 50,000 users. For evaluation we use the MIND test set, which is the same for both MIND-Small as for MIND-Large. 
The second dataset we use is Adressa \citep{gulla2017adressa}, which is sourced from traffic on the Adresseavisen news website in Norway. 
As in \citet{yi2023efficientfedrec}, we build data splits using one week of data. Our training split uses the first six days of click data, where the first five days serve as historical clicks and the sixth day provides impression and click data. The validation split consists of a random 20$\%$ sample from the last day, with the remaining 80$\%$ used for the test split. Since Adressa lacks impressions with negative samples, we randomly select 20 news articles to construct our impression data. See Table~\ref{tab:data_stats} in the appendix for summary statistics.

We compare our approach to baselines from the literature, including BERT-based news recommendation approaches from \citet{wu2021empowering}, which are constructed by replacing traditional text-encoding methods with BERT \citep{devlin-etal-2019-bert}. These methods include BERT-NPA based on \citet{wu2019npa}, BERT-LSTUR based on \citet{an-etal-2019-neural}, and BERT-NRMS based on \citet{wu-etal-2019-neural-news}. We furthermore compare to DIGAT \citep{mao-etal-2022-digat}, HDNR \citep{10.1145/3539618.3591693}, Prompt4NR \citep{Zhang_2023}, and UniTrec \citep{mao-etal-2023-unitrec}. For Prompt4NR, we only compare the performance with single model based variants. The ensembling version of Prompt4NR is not considered in this experiment for a fair comparison among approaches. In line with previous work we ignore baseline methods that use auxiliary inputs such as a topic label.

\begin{table*}[tphb]
    \centering
    \footnotesize
    \begin{tabular}{llccccc}
        \toprule
        \textbf{Model} & \textbf{Inference Strategy} & \textbf{AUC} & \textbf{MRR} & \textbf{nDCG@5} & \textbf{nDCG@10} & \textbf{Complexity} \\
        \midrule
        Pairwise & Pairwise & 65.99 &	30.76 &	33.64	& 39.67 & $O(n^2)$\\
        GLIMPSE & Pointwise & 67.90 &	31.65 & 34.92 & 41.59 & $O(n)$ \\
        GLIMPSE & Box filling & 67.98 & 32.05 & 35.13 &	41.60 & $O(n^2)$\\
        GLIMPSE & BubbleSort (random init) & 65.03 &	31.88 &	34.75 &	41.58 & $O(n^2)$ \\
        GLIMPSE & BubbleSort (point init) & 68.08 & 32.13 &	35.04 & 41.77 & $O(n^2)$ \\
        GLIMPSE & N-Window & 67.68 & 32.22 & 35.10 & 	41.86 & $O(m n)$\\
        GLIMPSE & S-window & 67.23 & 31.72 &	34.85 &	41.40 & $O(m n)$\\
        GLIMPSE & 1 RTL pass on top-3 & 68.54 & 32.41	& 35.48 &	42.16 & $O(n)$\\
        GLIMPSE & 1 RTL pass on top-5 & \textbf{68.59} &	\textbf{32.60} &	\textbf{35.60} &	\textbf{42.27} & $O(n)$\\
        GLIMPSE & 1 RTL pass on top-10 & 68.32	& 32.19 &	35.18 &	41.96 & $O(n)$\\
        GLIMPSE & 2 RTL passes on top-5 & 68.40 &	32.34 &	35.43 &	42.11 & $O(n)$\\
        \bottomrule
    \end{tabular}
    \caption{Comparison of different inference strategies on 10\% of MIND test set. For the N-Window and S-window strategies the notation corresponds to \citet{Gienapp_2022}. \label{tab:inf}}
\end{table*}

For GLIMPSE we use an encoder-decoder Flan-T5 model \citep{chung2022scaling} from the HuggingFace library\footnote{https://huggingface.co/} and fine-tune the model using the mixture-of-tasks approach for 4 epochs. We use a learning rate of $10^{-5}$ with a linear scheduler and perform early stopping. We use the sum of the pointwise and pairwise validation accuracies to select the best checkpoint. The results reported below are using the \textit{base} version of Flan-T5 which consists of 200 million parameters. During inference, we save the pointwise model along with its scores to subsequently perform the RTL pass. We use the same hyperparameters for both datasets and measure model performance.
\subsection{Results}
Table~\ref{tab:combined_results} reports the results  
for existing methods and several variants of GLIMPSE.
We show the performance of the \GLIMPSEpoint method (which refers to our model 
when trained in a multi-task fashion but without the RTL reranking pass 
applied during inference), as well as for GLIMPSE with the RTL pass applied on 
the top $k$ items in the pointwise-ranked list.
From the results on the MIND dataset, we can see that our best performing 
strategy improves over the strongest 
baseline, UniTRec, by 0.80\% on AUC, 0.36\% on MRR, 0.08\% on nDCG@5 and 0.32\% 
on nDCG@10.
For Adressa the improvements over baseline methods are even greater, with GLIMPSE improving on the strongest baseline by 7.95\% on AUC, 5.27\% on nDCG@5, and 6.59\% on nDCG@10, while matching performance on MRR. These results emphasize the benefit and efficiency of pairwise comparisons in our inference strategy, with performance improvements against pointwise inference resulting from only 8 pairwise comparisons on MIND and only 1 on Adressa.

\subsection{Inference Strategies}

Numerous inference strategies can be used to combine the predictions of a pointwise initial ranking and a pairwise comparison model. We analyze several in terms of performance and computational complexity. To evaluate inference strategies empirically we use a subset of 10\% of the MIND test set to allow comparison
to strategies that are quadratic in the number of items. We include the Prompt4NR method \citep{Zhang_2023} for reference.

Table~\ref{tab:inf} shows the results of these experiments. We use the multi-task GLIMPSE model as a baseline and only vary the inference strategy. The ``pointwise'' inference strategy refers to only using the relevance prediction for an item (see Section~\ref{sec:multitask}). Next, the ``box filling'' approach relies on performing all-pair comparisons to fill a table with pairwise preference scores, and ranking the items based on the marginals of this table.  This strategy simulates the method by \citet{jiang-etal-2023-llm}. We also compare to the traditional Bubblesort algorithm \citep{friend1956sorting}, using both random and pointwise-inference initializations.
We also compare to the ``neighborhood-window'' (N-Window) and ``skip-window'' (S-Window) methods proposed in \citet{Gienapp_2022}. These methods use a moving window of items as batches for pairwise reranking. Finally, we compare to four variants of our approach, using either one or two RTL passes on the top-$k$ items in the list ranked by pointwise scoring. We observe that the GLIMPSE approach with one RTL pass on the top-5 items ranked by pointwise scoring outperforms the other inference strategies. See Appendix~\ref{app:runtime} for additional results on running time.


\begin{figure*}
    \centering
    \resizebox{\textwidth}{!}{
    \begin{tabular}{ccc}
	\includegraphics[width=0.25\textwidth]{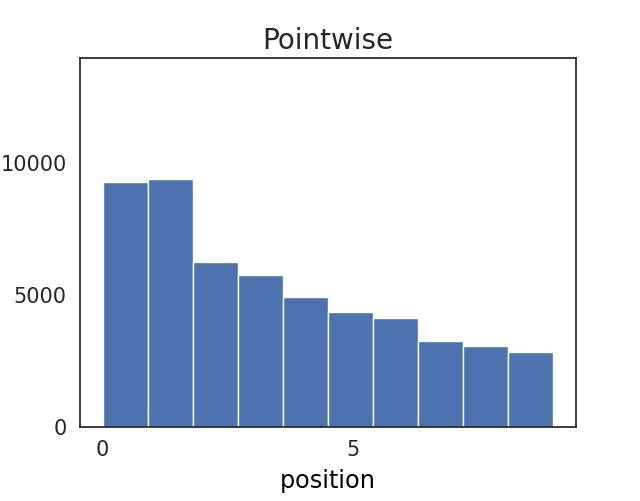} &
	\includegraphics[width=0.25\textwidth]{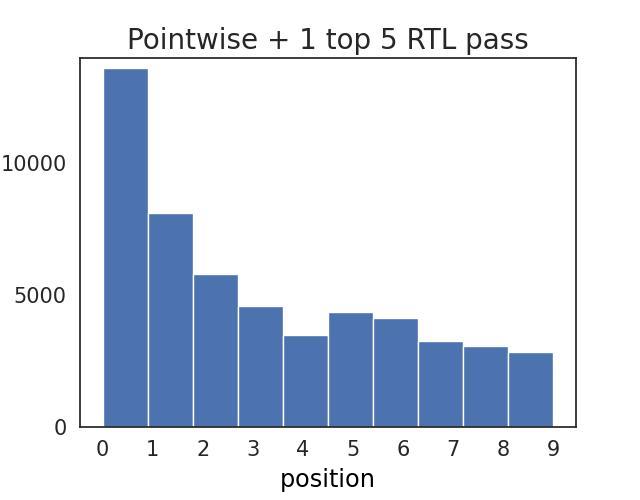} &
	\includegraphics[width=0.25\textwidth]{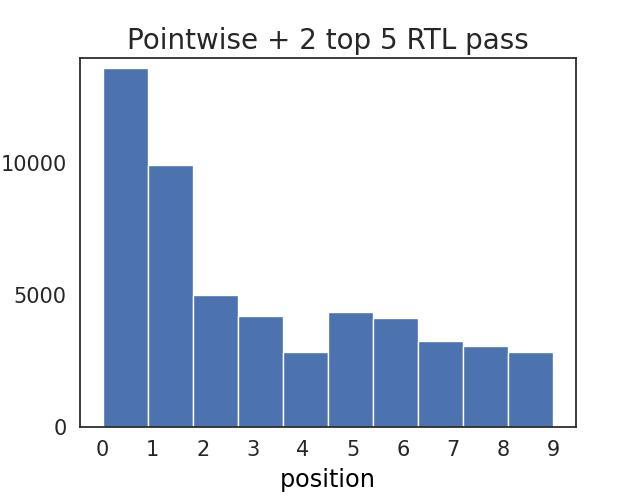} \\
	\end{tabular}
	}
	\caption{Comparison of the distribution shift after RTL passes. The figures show how often the first positive label is at each position. The figure on the left shows the distribution obtained from a pointwise model using 5\% of the data (see main text), the figure in the middle is based on the same weak pointwise model, but uses an RTL pass on the top-5 items. It can be seen that RTL passes progressively move the first positive label to the front of the ranked list. 
	\label{fig:ablate}}
\end{figure*}

\subsection{Ablation Study}

To study how our framework performs with varied strengths of the pointwise 
model we train a weak pointwise model with a subset of pointwise datapoints and 
subsequently perform RTL passes using a strong pairwise model.
For the weak pointwise model, we retain 5\% and 10\% respectively of the 
pointwise datapoints and use the weak model for the initial pointwise ranking.
We then use a strong pairwise model trained on the complete pairwise data for 
RTL passes.
The results of this experiment are reported in Table~\ref{tab:ablate_table}.
This shows that with a 5\% pointwise model, performing a single RTL pass 
results in a relative increase of 2.96\% on AUC, 12.91\% on MRR, 9.60\% on 
nDCG@5 and 7.85\% on nDCG@10, while performing two RTL passes results in an 
increase of 3.59\% on AUC, 14.74\% on MRR, 11.06\% on nDCG@5, and 9.01\% on 
nDCG@10.
On the 10\% pointwise model a single pass shows relative improvement of 1.70\% 
on AUC, 5.02\% on MRR, 5.43\% on nDCG@5, and 5.54\% on nDCG@10.
Thus we observe that the performance gains from 
pairwise reranking increase as the pointwise model gets weaker.
Note that the gains obtained from RTL passes in this experiment with weak 
pointwise models are much higher than those in 
Table~\ref{tab:combined_results}, confirming the above observation.
Figure~\ref{fig:ablate} shows the frequency distribution of the first 
occurrence of positive label in the predicted rankings at different inference 
stages. After applying RTL passes, we see a shift in the histogram towards the left side, indicating the positive labels are being pushed towards the top positions. Finally, to establish the benefit of multi-task training for 
pointwise ranking, we compare the performance of pointwise ranking with and 
without the pairwise comparison task.
The results in Table~\ref{tab:multi_ablate_table} show that without the 
multi-task training there is a significant performance drop, 
especially in the ranking metrics MRR ($-1.52\%$),  nDCG@5 
($-2.17\%$) and nDCG@10 ($-1.31\%$).
This confirms our argument that the task of recommendation 
naturally benefits from comparisons between candidates. 


\begin{table*}
    \centering
    \footnotesize
    \begin{tabular}{ccccc}
        \toprule
        \textbf{Model} & \textbf{AUC} & \textbf{MRR} & \textbf{nDCG@5} & \textbf{nDCG@10} \\
        \midrule
        \GLIMPSEpoint{} (5\%) & 56.99 &	25.71	& 27.48	& 33.60\\
       w/ pair (1pass, top5) & 58.68 &	29.03 &	30.12 &	36.24 \\
       w/ pair (2pass, top5) & 59.04	& 29.50 &	30.52 & 	36.63 \\
    \midrule
      \GLIMPSEpoint{} (10\%) & 65.61 & 31.02 &	34.05 &  40.38 \\
      w/ pair (1pass, top5) & 66.73	 & 32.58 &	35.90 &	42.62 \\
        \bottomrule
    \end{tabular}
    \caption{Ablation study comparing a weak pointwise model (trained on a 5\% or 10\% subset of pointwise data) and a strong pairwise model (trained on all pairwise data), using the MIND dataset. \label{tab:ablate_table}}
\end{table*}

\begin{table*}
    \centering
    \small
    \begin{tabular}{ccccc}
        \toprule 
        \textbf{Model} & \textbf{AUC} & \textbf{MRR} & \textbf{nDCG@5} & \textbf{nDCG@10} \\
        \midrule
        \GLIMPSEpoint & 68.18 &	33.56 &	37.27 & 43.48 \\
        pointwise & 68.63 &	33.05 &	36.56 &	42.91 \\
        \bottomrule
    \end{tabular}
    \caption{Results of an ablation study to illustrate the benefit of multi-task training for pointwise ranking. The effect of multi-task training is especially noticeable in the ranking metrics. \label{tab:multi_ablate_table}}
\end{table*}

\section{Related Work} 
\label{sec:related_work}
In this section, we focus primarily on news recommendation approaches that leverage LLMs, as well as the pairwise ranking literature.

\subsection{LLM-based News Recommendation}

Prompt4NR \citep{Zhang_2023} is a recent baseline for news recommendation on the MIND dataset. The paper proposes continuous and discrete prompts to provide history and treat the task of click prediction as a mask-filling task. \citet{10.1145/3539618.3592003} propose a multi-task pre-training approach for news recommendation. \citet{10.1145/3539618.3591753} propose RCENR, an explainable model that generates user or news subgraphs to enhance recommendation and extend the dimensions and diversity of reasoning.  \citet{10.1145/3539618.3591693} introduce HDNR, a hyperboloid model with exponential growth capacity to conduct user and news modeling. UniTrec \citep{mao-etal-2023-unitrec} is a unified generative framework for three text-based recommendation tasks, including news recommendation. It leverages candidate perplexity and discriminative scores to perform final pointwise ranking. \citet{yu2022tinynewsrec} propose Tiny-Newsrec, a self-supervised domain-specific post-training method to address the domain shift problem from pre-training tasks to downstream news recommendation.
DebiasGAN \citep{wu-etal-2022-debiasgan} alleviates position bias via adversarial learning by modelling the personalized effect of position bias on click behavior to estimate unbiased click scores. DIGAT \citep{mao-etal-2022-digat} is a dual-interactive graph attention network to model user and news graph channels. \citet{li-etal-2022-miner} introduce MINER, a poly-attention scheme to learn multiple interest vectors for each user to effectively model different aspects of user interest. MTRec \citep{bi-etal-2022-mtrec} is a multi-task method to incorporate the multi-field information in order to enhance the news encoding capabilities. \citet{wu-etal-2022-two} propose UniRec, a unified method for recall and ranking in news recommendation. \citet{gong2022positive} propose a framework that leverages both positive and negative feedback. DRPN \citep{hu2022denoising} de-noises both positive and negative implicit feedback to simulate noisy real world use-cases. 
\subsection{Pairwise Comparisons for Ranking}

Pairwise ranking has been a long-standing approach ever since the method of ranking  using paired comparisons was introduced by \citet{Bradley1952RankAO}. RankNet \citep{10.1145/1102351.1102363} is a gradient descent method for pairwise ranking. LambdaRank \citep{10.5555/2976456.2976481} builds on RankNet by incorporating IR measures (e.g., nDCG) into the derivative of the cost function. LambdaRankMart \citep{Burges2010FromRT} is a boosted version of LambdaRank. PiRank \citep{swezey2021pirank} adds new class of differentiable surrogate functions for ranking.
\citet{pmlr-v84-heckel18a} propose an active learning approach to select pairs for comparisons that results in an approximate rank with logarithmic complexity.
\citet{Gienapp_2022} propose to sparsify the number of pairwise comparisons using next M window and skip K window strategies. This way the number of comparisons is less than quadratic and hence achieve a middle ground. \citet{qin2023large} propose \textit{Pairwise Prompting} to perform ranking by sorting items from pairwise comparisons and compare listwise and pairwise strategies and show that bubble sort with pairwise comparisons achieves best results. \citet{hou2023large} formalize the recommendation problem as a conditional ranking task and adopt recency-focused in-context prompting and candidate generation algorithms before directly performing listwise ranking.  \citet{pradeep2021expandomonoduo} address the pipelined recommendation problem with a candidate retriever followed by a 'mono' step where they rank using pointwise probability and a subsequent 'duo' step where they use pairwise comparisons for all-pairs. They test different aggregation functions for the all-pairs pairwise scores to re-rank the candidates. LLM Blender \citep{jiang-etal-2023-llm} is a box-filling strategy with a pairwise ranker for aggregating pairwise scores for ranking. \citet{Dai_2023} use listwise, pairwise, and pointwise prompting on ChatGPT by choosing a fixed candidate set for experiments. \citet{liusie2023llm} use pairwise comparative assessment for NLG evaluation.

\section{Conclusion}
We proposed GLIMPSE, an algorithm for personalized recommendation that leverages PLMs alongside a novel inference strategy to efficiently combine pointwise relevance prediction and pairwise comparison. We presented a multi-task finetuning approach to facilitate this inference and conducted extensive experiments on real-world datasets, demonstrating that GLIMPSE outperforms state-of-the-art methods on the MIND and Adressa datasets. Moreover, we provided a rigorous theoretical analysis of our proposed approach and derived conditions under which the RTL re-ranking pass is beneficial. Our work underscores the potential of leveraging language models for news recommendation, emphasizing both performance and real-world efficiency.

\section{Limitations}
The proposed algorithm is proved to achieve better performance than pointwise ranking approaches. However, it's important to note that this improvement may be modest when dealing with an extensive dataset used for training the pointwise relevance prediction model. If the pointwise model achieves sufficient accuracy, incorporating a pairwise preference model through the RTL rank aggregation strategy might offer limited additional value. Our theoretical findings support this observation, indicating that the gains could even turn negative should the state distribution derived from pointwise predictions fail to meet the defined conditions. While GLIMPSE generally improves ranking performance, its effectiveness hinges on both the size of the training set and the precision of the pointwise relevance prediction model.

\bibliography{custom}

\appendix

\appendix
\onecolumn
\section{Appendix}
\subsection{Proof of Theorem~\ref{the:main}}
\label{app:proof}
In this section, we present the proof of Theorem~\ref{the:main}. Recall that in our analysis we use Markov chain theory to analyze the outcomes from both the point-wise and pairwise stages within the inference strategy. We construct a Markov chain on a discrete state space encompassing all permutations of the ranking results. We define an initial probability distribution over the state space based on the stochastic ranking results generated by the pointwise inference stage. Through pairwise refinement the distribution over states shifts as stochastic pairwise swaps are applied by the pairwise inference stage. Thus, pairwise inference provides a transition probability kernel over the state space. By Theorem~\ref{the:main} we have that if the induced transition probability matrix satisfies certain conditions, we are guaranteed to achieve positive gain from two-stage inference. Below we present the proof of this theorem.

The gain obtained after our two-stage inference process compared to the outcome directly from the pointwise inference can be written as
\begin{equation}
\label{Eq:expectedGain}
\begin{array}{lll}
    G&=&\mathbb{E}_{\mathcal{A}}(\Delta)-
    \mathbb{E}_{\boldsymbol{\pi}}(\Delta)= \sum_{\mathbf{z}'} \boldsymbol{\pi} \mathbf{T}_{\mathbf{z}'} \Delta(\mathbf{z}' \given \mathcal{H})\\
    &&-\sum_{\mathbf{z}'}\boldsymbol{\pi}\Delta(\boldsymbol{z}' \given \mathcal{H})\\ &=& \boldsymbol{\pi}(\mathbf{T}-\mathbf{I})\Delta^T,
\end{array}
\end{equation}
where $\Delta = [\Delta(\mathbf{z}_1 \given \mathcal{H}),\Delta(\mathbf{z}_2 \given \mathcal{H}),...,\Delta(\mathbf{z}_K \given \mathcal{H})]^T$ and $\mathbf{I}$ is the identity matrix. Without loss of generality, we only provide the proof with MRR as the ranking metric for simplicity, i.e., $\Delta=[1,\frac{1}{2},...,\frac{1}{K}]^T$. Note that the proof in this section applies to any non-increasing ranking metric.

In order to analyze the gain defined above, we re-write the matrix  $\mathbf{T}-\mathbf{I}$ by plugging in the definition of $\mathbf{T}$ shown in Eq.~\ref{Equation:transitionMatrix} as
\begin{equation*}
\left[
    \begin{array}{llll}
        -\mu & (1-\mu)\nu & (1-\mu)\nu^2 & ... \\
        \mu  & -\mu-\nu+\mu\nu & (1-\mu)(1-\nu)\nu & ... \\
        0 & \mu & -\mu-\nu+\mu\nu & ...  \\
        \vdots & \vdots & \vdots & ... \\
        0 & 0 & 0 & ... 
    \end{array}\right.
\end{equation*}
\begin{equation}
\left.
    \begin{array}{ll}
    (1-\mu)\nu^{K-2} & \nu^{K-1} \\
    (1-\mu)(1-\nu)\nu^{K-3} & (1-\nu)\nu^{K-2}\\
    (1-\mu)(1-\nu)\nu^{K-4} & (1-\nu)\nu^{K-3}\\
    \vdots & \vdots\\
    \mu & -\nu
    \end{array}
\right].
\end{equation}
The matrix $\mathbf{T}-\mathbf{I}$ is an upper Hessenberg matrix whose diagonal elements are negative. The sub-diagonal elements are always equal to $\mu$. As the ranking metric is a non-increasing function, $\mathbf{T}-\mathbf{I}$ will lead to positive gain when the diagonals  and sub-diagonal are small enough.  Formally, we can rewrite Eq.~\ref{Eq:expectedGain} by following element-wise matrix multiplication as
\begin{equation*}
    \begin{array}{lll}
    G &=& -\mu \pi_1/2 +\sum_{i=2}^{K-1}\pi_{i}\left[(1-\mu)\nu^{(i-1)}\right.\\
    &&+\left.(1-\mu)(1-\nu)\sum_{j=1}^{i-2}\frac{\nu^j}{i-j}\right. \\
    &&+\mu/(i+1)+ \left.(\mu\nu-\mu-\nu)/i\right]\\
    &&+\pi_{K}\left(\nu^{K-1}-\nu/K\right.\\
    &&\left.+(1-\nu)\sum_{i=2}^{K-1}\nu^{K-i}/i\right). 
    \end{array}
\end{equation*}
It is clear that $G$ is a bi-variate $(K-1)$-th degree polynomial function in terms of $\mu$ and $\nu$.  We decompose $G$ into $K-1$ components $G=G_1+G_2+...+G_{K-1}$ whose $k$th component is the sum of all order $k$ terms in $G$. 
We start with the sum of all first order terms $G_1$, 
\begin{equation*}
    \begin{array}{lll}
        G_1&=&-\frac{\pi_1\mu}{2}+\pi_2\left(\nu+\frac{\mu}{3}-\frac{\mu+\nu}{2}\right)\\
        &&+\sum_{i=3}^{K-1}\pi_i\left(\frac{\nu}{i-1}+\frac{\mu}{i+1}-\frac{\mu+\nu}{i}\right)\\
        &&-\frac{\pi_K\nu}{K}+\frac{\pi_K\nu}{K-1}   \\
        &=&\sum_{i=1}^{K-1}\left(\frac{\pi_{i+1}\nu}{i(i+1)}-\frac{\pi_{i}\mu}{i(i+1)}\right).
    \end{array}
\end{equation*}
The sum of all first order terms is guaranteed to be positive if we have $\mu\leq\frac{\pi_{i+1}}{\pi_i}\nu$, for all $i=1,...,K-1$. Similarly, the sum of all the second order terms can be written as
\begin{equation*}
\begin{array}{lll}
     G_2&=& -\pi_2\frac{\mu\nu}{2}+
     \frac{\pi_K\nu^2}{K-2}-\frac{\pi_K\nu^{2}}{K-1}\\
     &&+\sum_{i=3}^{K-1}\pi_i\left(\frac{\nu^2}{i-2}-\frac{\mu\nu}{i-1}-\frac{\nu^2}{i-1}+\frac{\mu\nu}{i}\right)\\
     &=& \sum_{i=2}^{K-1}\left(\frac{\pi_{i+1}\nu^2}{i(i-1)}-\frac{\pi_{i}\mu\nu}{i(i-1)}\right). 
\end{array}
\end{equation*}
We have $G_2\geq0$ if $\mu\leq\frac{\pi_{i+1}}{\pi_{i}}\nu, \forall i=2,...,K-1$. For any $k$-th order terms, there is
\begin{equation*}
    \begin{array}{lll}
         G_k&=&\sum_{i=k}^{K-1}\left(\frac{\pi_i\nu^{k}}{(i-k-1)(i-k)}-\frac{\pi_i\mu\nu^{k-1}}{(i-k)(i-k+1)}\right).
    \end{array}
\end{equation*}
For the last term with order $K-1$,
\begin{equation*}
\begin{array}{lll}
   G_{K-1} &= &-\pi_{K-1}\mu\nu^{K-2}+\pi_{K-1}\frac{\mu\nu^{K-2}}{2}\\
   &&+\pi_K\nu^{K-1}
   -\frac{\pi_K\nu^{K-1}}{2} \\
   &&= \frac{\pi_K\nu^{K-1}}{2}-\frac{\pi_{K-1}\mu\nu^{K-2}}{2}.
\end{array}
\end{equation*}
When $\mu$ and $\nu$  satisfy $\mu\leq\frac{\pi_i}{\pi_{i+1}}\nu,\forall i=1,...,K-1$, we are guaranteed to have positive gain $G$ from one pass of RTL inference.

\subsection{Experiment Details}
\label{app:expdetails}
In this section, we provides more detailed description for the experimental setup used in Section~\ref{sec:experiments}. 

\subsubsection{Datasets}
The statistics for datasets used in our experiment in shown in Table~\ref{tab:data_stats}.

\subsubsection{Model architecture and fine-tuning}

The LLM chosen in this paper is the Flan-T5-base model \citep{chung2022scaling}, obtained from HuggingFace. We chose the base version, which has a parameter size of 200M, in order to maintain the same model size as previous works for a fair comparison. We use the Flan version of T5, which is additionally instruction-tuned, as we employ instruction prompts for our multi-task training. 

We finetuned the Flan-T5 model in a multi-task setup as described in Section~\ref{sec:multitask}. The model is trained by maximizing the likelihood objective given in Eq~\ref{eq:LED}. We train the model for 4 epochs using our mixture-of-tasks approach with a linear learning rate schedule starting from 1e-5 with early stopping. We use the validation split of the dataset for deciding hyperparameters used for training. 

\subsubsection{Prompt Construction}
For the MIND dataset, we use the same prompts used by \citet{Zhang_2023} to facilitate fair comparison. We utilize the headings of the news articles in MIND-small training dataset. To limit the input prompt length to 512 tokens, we include a maximum of 50 user click history items, each with a maximum length of 10 tokens. The entire history is capped at a maximum length of 450 tokens. Additionally, for each candidate title, we allow a maximum of 20 tokens. All these hyperparameters align with the code repository of \citet{Zhang_2023}. We use the T5 separator token ‘<s>' to separate history as well as candidate items. The instruction prompts used along with the context and candidates are reported in Table~\ref{tab:prompt_table}. For the QuoteRec dataset, we reuse the prompts reported by UniTrec \citep{mao-etal-2023-unitrec}. We utilize the same instruction prompts as for MIND, as well as the same item separators.

\subsection{Additional Experiments}
\label{sec:AdditionalExps}

In this section, we present additional experiments for analyzing the performance of the proposed approach.
\subsubsection{From pair-wise to list-wise}
To understand the performance of GLIMPSE beyond pair-wise, we delve further into extending the approach to list-wise input by incorporating more than two samples at a time during training. Without loss of generality, we consider triple-wise comparison task only in this experiment. More specifically, during training, we present the model with three candidate documents simultaneously and ask it to pick out the most relevant one. Employing a sampling strategy akin to the pairwise task outlined in Section~\ref{sec:multitask}, we ensure a balanced representation of training samples across point-wise, pair-wise, and triple-wise tasks. For evaluation, we compared the point-pair-triple wise multi-task trained model with GLIMPSE on point-wise ranking inference task. The result is presented in Table~\ref{tab:triplewise}.

Our results indicate that the inclusion of this triple-wise comparison task led to a degradation in pointwise performance in Table~\ref{tab:triplewise}. This shows that processing more than two candidate items at a time poses a challenge for small-sized models, which justifies our proposed combination of pointwise and pairwise approaches. Furthermore, this finding aligns with claims from related literature \citep{qin2023large, sun2023chatgpt} that show that list-wise approaches perform more poorly than pairwise ones.

\subsubsection{Effect of point-wise initialisation}
As we presented, GLIMPSE is a two-stage approach which the point-wise and pair-wise inferences are used for initialisation and refinement respectively. To understand the effect of point-wise initialisation, we conduct an experiment to compare random initialisation with the proposed point-wise initialization approach. More specifically, with the same multi-task trained model, we perform different inference strategies. The first strategy is based on random initialization, where we perform RTL passes on a randomly-initialized list instead of utilizing a point-wise prior on the MIND dataset. The result is presented in Table~\ref{tab:randomInitialisation}. We can observe that the results of random initialisation are much worse compared to the proposed point-wise initialisation followed by RTL passes. This observation holds even as we increase the number of RTL passes.

Beside empirical results, our main theory presented also provides the justification of the point-wise initialization. From a theoretical standpoint, the point-wise initialization yields a more advantageous initial state distribution in the underlying Markov chain. As demonstrated in the main paper, the initial distribution, or prior distribution resulting from point-wise inference, can be expressed as Eq.~\ref{Equ:prior}. The induced distribution can be approximated as an exponential distribution, with its parameter determined by the precision of the point-wise model. This implies that a robust point-wise inference model can offer an adequate starting distribution for our two-stage inference approach. In contrast, a randomly-initialized prior distribution would adhere to a uniform distribution, limiting the extent of enhancement that the pairwise inference can achieve via the transition matrix.

\subsubsection{Quote Recommendation Task}
In addition to the news recommendation datasets, we demonstrate that our approach is competitive on other text recommendation tasks. We use the QuoteRec dataset \citep{wang-etal-2020-continuity}, which focuses on recommending a quotation appropriate to a conversational context. For fair comparison to previous work, we use the \emph{Reddit-quotation} dataset with the same splits as in \citet{mao-etal-2023-unitrec}. Each conversation in this dataset has one positive label, with 1111 quotations in total. For this dataset we follow the work of \citet{mao-etal-2023-unitrec} to ensure we use the same splits, and thus compare to BERT4Rec \citep{10.1145/3357384.3357895}, RoBERTa-Sim \citep{qi-etal-2022-quoter}, UNBERT \citep{Zhang2021UNBERTUM}, and UniTRec \citep{mao-etal-2023-unitrec}. Summary statistics can be found in Table~\ref{tab:data_stats} and results on this dataset are in Table~\ref{tab:quoterec_results}. The performance of our approach on the QuoteRec dataset follows the same trend as above: pointwise scores are improved as we perform RTL passes. Here we see that our method outperforms all existing baselines except UniTRec 
\citep{mao-etal-2023-unitrec}. We hypothesize that this is because the QuoteRec dataset has only one positive and 1110 negative samples in each impression, which reduces the benefit of pairwise ranking.

\begin{table*}[t]
  \centering
  \footnotesize
  \begin{tabular}{llrrr}
    \toprule
    \textbf{Dataset} & \textbf{Property} & \textbf{Train}  & \textbf{Val.}  & \textbf{Test} \\
    \midrule
    MIND & Impressions & 149116 & 7849 & 73152 \\ 
    MIND & Users & 49123 & 6981 & 50000 \\
    Adressa & Impressions & 290523 & 63226 & 252902 \\
    Adressa & Users & 131740 & 46919 & 115458 \\
    QuoteRec & Conversations & 35633 & 4454 & 4454 \\
    QuoteRec & Quotes & 1111 & 830 & 795 \\
    \bottomrule
  \end{tabular}
  \caption{Summary statistics for datasets used in our experiments. In line with existing literature we use the MIND-Small subset of MIND for training, and the MIND test set for evaluation. \label{tab:data_stats}}
\end{table*}

\begin{table*}
  \centering
  \footnotesize
  \begin{tabular}{p{1.5cm}|p{7.5cm}|p{3.7cm}}
    \toprule
    Task & Input prompt  & Target word \\
    \midrule
    \textit{Rel(.)} & Given that user has clicked on $<$user\_history$>$. Is candidate A: $<$candidate\_title$>$ a good recommendation to user? Respond Yes or No. & Yes/No \\ 
    \midrule
    \textit{Pref(.)} & Given that user has clicked on $<$user\_history$>$, amongst the 2 news titles, candidate A: $<$candidate\_title1$>$ and candidate B: $<$candidate\_title2$>$ which is more appropriate for recommendation to user? Respond Candidate A or Candidate B. & Candidate A/ Candidate B\\
    \bottomrule
  \end{tabular}
  \caption{Sample prompts used for each task. \label{tab:prompt_table}}
\end{table*}
\begin{table*}[tb]
    \centering
    \footnotesize
    \begin{tabular}{lcccc}
        \toprule
        \textbf{Model} & \textbf{AUC} & \textbf{MRR} & \textbf{nDCG@5} & \textbf{nDCG@10} \\
        \midrule
        Multi-task Point-Pair-Triple & 68.17 &	33.13 &	36.65	& 42.84 \\
        Multi-task Point-Pair & 68.18 &	33.56 & 37.27 & 43.48 \\
        \bottomrule
    \end{tabular}
    \caption{Comparison of triple-wise and pair-wise on MIND test set.}
    \label{tab:triplewise}
\end{table*}

\begin{table*}
    \centering
    \footnotesize
    \begin{tabular}{lcccc}
        \toprule
        \textbf{Initialisation Method} & \textbf{AUC} & \textbf{MRR} & \textbf{nDCG@5} & \textbf{nDCG@10} \\
        \midrule
        Random & 52.14 &	25.48 &	25.31	& 31.65 \\
        Point-wise inference & 68.97 &	33.73 & 37.41 & 43.62 \\
        \bottomrule
    \end{tabular}
    \caption{Comparison of different inference initialisation methods on MIND test set.}
    \label{tab:randomInitialisation}
\end{table*}

\begin{table*} 
\centering
\footnotesize
\begin{tabular}{lccc}
\toprule
\textbf{Model} & \textbf{MRR} & \textbf{nDCG@5} & \textbf{nDCG@10} \\
\midrule
Bert4Rec & 33.59 & 34.26 & 37.37 \\
RoBERTa-Sim & 37.13	& 37.96 & 41.18 \\
UNBERT &  39.75 & 40.74 & 43.69 \\
UniTRec & \textbf{41.24} & \textbf{42.38} & \textbf{45.31} \\
\midrule
\GLIMPSEpoint & 40.18 & 41.39 & 43.70 \\
\; w/ pair (top 2) & 40.31 & 41.46 & 43.77 \\
\; w/ pair (top 5) & 40.43 & 41.53 & 43.85 \\
\bottomrule
\end{tabular}
\caption{Results on the QuoteRec dataset compared to baseline methods. Results for baseline methods are extracted from the respective papers. Inference with GLIMPSE is performed using a single RTL pass. \label{tab:quoterec_results}}
\end{table*}

\subsubsection{Runtime Comparison}
\label{app:runtime}
In order to show the efficiency of the proposed GLIMPSE algorithm, we also compare the inference times with baselines in Table~\ref{tab:inf}. The result is shown in Table~\ref{tab:runtime}. We include the results to present the runtime of our approach along with other inference baselines. We report the running time results, measured in seconds, for our approach and baseline inference methods on 10\% of the MIND test data in the following table. These results are based on three repeated runs of each experiment, with the average and standard deviation provided to ensure reliability. For some baselines, we were unable to obtain a complete runtime due to excessive duration. In these cases, we estimated the runtime based on early stopping at 10\% completion. 
\begin{table*}
\centering
\scriptsize
\begin{tabular}{lcccccccc}
\toprule
\textbf{Inference Method} & \textbf{Pairwise} & \textbf{Pointwise} & \textbf{Boxfilling} & \textbf{BubbleSort} &
\textbf{N-window} & \textbf{S-window} & \textbf{1 RTL top-5} & \textbf{1 RTL top-10} \\
\midrule
Runtime (s) & $>20000$ & 651.9$\pm$0.8 & $>$20000 & $>$20000 & $>$8000 & $>$9000 & 1234.1$\pm$7.5 & 1872.1$\pm$15.1\\
\bottomrule
\end{tabular}
\caption{Inference runtime comparison (in seconds) of different approaches on 10\% of MIND test data.}
\label{tab:runtime}
\end{table*}

\subsubsection{Model Size Considerations for News Recommendation}
While large language models (LLMs) have demonstrated impressive capabilities in recommendation tasks \citep{liu2023chatgpt, qin2023large} , news recommendation presents unique challenges that necessitate careful consideration of model size. The fast-paced nature of news, and an evolving information landscape, demands rapid response times and efficient computation. Consequently, existing body of research in news recommendation primarily focuses on smaller, more efficient models \citep{Zhang_2023, mao-etal-2023-unitrec}. In line with this emphasis on practicality, our work prioritizes model size considerations to ensure real-world applicability. Our proposed integration of point-wise and pairwise learning with $O(K)$ complexity further exemplifies how news ranking performance can be enhanced while maintaining efficiency, a crucial aspect in news recommendation.

\end{document}